\documentclass[conference]{IEEEtran}
\IEEEoverridecommandlockouts

\usepackage[utf8]{inputenc} 
\usepackage[hidelinks]{hyperref} 

\usepackage{cite}
\usepackage{tikz}
\usetikzlibrary{fit}
\usepackage{tikz}
\usetikzlibrary{shapes.geometric, arrows, fit}

\usetikzlibrary{shapes.geometric, arrows}
\usepackage{amsmath,amssymb,amsfonts}
\usepackage{algorithmic}
\usetikzlibrary{positioning}
\usepackage{graphicx}
\usepackage{textcomp}
\usepackage{xcolor}
\usepackage{booktabs}
\usepackage{tikz}
\usetikzlibrary{shapes.geometric, arrows, fit, positioning, backgrounds}
\usepackage{float}

\usepackage{multirow}
\usepackage{array}

\usepackage{booktabs}
\usepackage[table]{xcolor}
\usepackage{array}
\usepackage{caption}

\definecolor{tableblue}{HTML}{2E74B5}

\def\BibTeX{{\rm B\kern-.05em{\sc i\kern-.025em b}\kern-.08em
    T\kern-.1667em\lower.7ex\hbox{E}\kern-.125emX}}
\begin{document}

\title{\vspace{0.2in} A Multi-Agent LLM Defense Pipeline Against Prompt Injection Attacks\\
}

\author{
S M Asif Hossain\textsuperscript{1}, 
Ruksat Khan Shayoni\textsuperscript{1}, 
Mohd Ruhul Ameen\textsuperscript{2}, 
Akif Islam\textsuperscript{3}, 
M. F. Mridha\textsuperscript{4}, 
Jungpil Shin\textsuperscript{5} \\[0.6em]

\textsuperscript{1}\textit{School of Computing, Wichita State University, Kansas, USA} \\
Emails: sxhossain10@shockers.wichita.edu, rxshayoni@shockers.wichita.edu\\[0.3em]

\textsuperscript{2}\textit{College of Engineering and Computer Sciences, Marshall University, Huntington, WV, USA} \\
Email: ameen@marshall.edu \\[0.3em]

\textsuperscript{3}\textit{Department of Computer Science and Engineering, University of Rajshahi, Bangladesh} \\
Email: s1910776135@ru.ac.bd \\[0.3em]

\textsuperscript{4}\textit{Department of Computer Science and Engineering, American International University-Bangladesh, Dhaka, Bangladesh} \\
Email: firoz.mridha@aiub.edu \\[0.3em]

\textsuperscript{5}\textit{School of Computer Science and Engineering, The University of Aizu, Aizuwakamatsu, Japan} \\
Email: jpshin@u-aizu.ac.jp \\[0.3em]
}

\maketitle

\begin{abstract}
Prompt injection attacks represent a major vulnerability in Large Language Model (LLM) deployments, where malicious instructions embedded in user inputs can override system prompts and induce unintended behaviors. This paper presents a novel multi-agent defense framework that employs specialized LLM agents in coordinated pipelines to detect and neutralize prompt injection attacks in real-time. We evaluate our approach using two distinct architectures: a sequential chain-of-agents pipeline and a hierarchical coordinator-based system. Our comprehensive evaluation on 55 unique prompt injection attacks, grouped into 8 categories and totaling 400 attack instances across two LLM platforms (ChatGLM and Llama2), demonstrates significant security improvements. Without defense mechanisms, baseline Attack Success Rates (ASR) reached 30\% for ChatGLM and 20\% for Llama2. Our multi-agent pipeline achieved 100\% mitigation, reducing ASR to 0\% across all tested scenarios. The framework demonstrates robustness across multiple attack categories including direct overrides, code execution attempts, data exfiltration, and obfuscation techniques, while maintaining system functionality for legitimate queries.
\end{abstract}


\section{Introduction}

Large Language Models (LLMs) have become integral components of modern applications, powering chatbots, code assistants, and automated decision systems \cite{radford2019language, brown2020language}. However, their widespread adoption has introduced novel security vulnerabilities, particularly the prompt injection attacks, where adversarial inputs can manipulate model behavior by overriding system instructions \cite{liu2023formalizing, li2024gentel}. The OWASP Top 10 for LLM Applications identifies prompt injection as the primary security risk \cite{owasp2023top10} which highlighting the urgent need for robust defense mechanisms.

Traditional security approaches, including static input sanitization and content filtering, prove inadequate against sophisticated prompt injection techniques \cite{greshake2023not, robey2023smoothllm}. These attacks exploit the fundamental architecture of LLMs, where system prompts and user inputs are processed as unified text sequences, enabling malicious instructions to override intended behaviors \cite{liu2023prompt}. Recent research indicates that even well-trained models with safety alignment remain vulnerable to carefully crafted adversarial prompts \cite{carlini2023aligned, wei2023jailbroken}.

Existing defense strategies fall into several categories: input preprocessing \cite{kumar2023certifying}, output filtering \cite{zhang2023defending}, prompt engineering \cite{wallace2019universal}, and model fine-tuning \cite{ziegler2019fine}. However, these approaches often exhibit limitations in handling novel attack vectors and maintaining system utility. Multi-agent architectures offers a promising alternative by utilizing distributed intelligence to implement defense-in-depth strategies \cite{wang2023self, jiang2023selfdefend}.

This paper introduces a comprehensive multi-agent defense pipeline that addresses prompt injection vulnerabilities through coordinated LLM agents. Our contributions include:

\begin{enumerate}
    \item \textbf{Novel Architecture Design}: Two complementary multi-agent configurations providing flexible deployment options for different security requirements.
    \item \textbf{Comprehensive Evaluation Framework}: Systematic assessment using 55 unique prompt injection attacks, grouped into 8 categories and totaling 400 attacks across two LLM platforms.
    \item \textbf{Empirical Validation}: Demonstration of 100\% attack mitigation across all tested scenarios while preserving system functionality.
    \item \textbf{Practical Implementation Guidelines}: Detailed analysis of deployment considerations, performance trade-offs, and scalability factors.
\end{enumerate}

\section{Related Work}

\subsection{Prompt Injection Attack Taxonomy}

Prompt injection attacks have been systematically categorized by Liu et al. \cite{liu2023formalizing}, who identify direct injection (explicit instruction override) and indirect injection (malicious content in external sources) as primary vectors. Recent work by Wang et al. \cite{wang2024protect} extends this taxonomy to include advanced obfuscation techniques and multi-turn persistent attacks.

\subsection{Existing Defense Mechanisms}

Current defense approaches can be classified into four main categories:

\textbf{Input Sanitization}: Traditional approaches employ rule-based filtering and keyword detection \cite{russinovich2024great}. However, these methods struggle with obfuscated or semantically disguised attacks \cite{zou2023universal}.

\textbf{Output Monitoring}: Post-generation filtering attempts to detect malicious content in model outputs \cite{li2023multi}. While effective for obvious violations, subtle attacks may evade detection \cite{zheng2024prompt}.

\textbf{Prompt Engineering}: Techniques such as instruction hierarchy and defensive prompting aim to make system prompts more resistant to override attempts \cite{deng2021attentionviz, anil2022constitutional}. The Polymorphic Prompt Assembly (PPA) approach by Wang et al. \cite{wang2024protect} randomizes prompt structure to prevent predictable attacks.

\textbf{Model-Level Defenses}: Approaches including adversarial training and reinforcement learning from human feedback (RLHF) aim to improve inherent model robustness \cite{ouyang2022training, bai2022constitutional}.

\subsection{Multi-Agent Security Frameworks}

Recent research explores multi-agent architectures for AI safety. Muliarevych \cite{muliarevych2024enhancing} proposed a layered defense system with specialized analyzer and validator agents. Gosmar et al. \cite{gosmar2025multi} outline multi-agent frameworks with generator, sanitizer, and policy enforcer components. Our work extends these concepts by implementing comprehensive multi-agent pipelines with empirical validation across diverse attack scenarios, and our analysis shows that the approach is especially effective against high-risk categories such as delegate/tool-manipulation attacks, role-play coercion, reconnaissance/environment leakage, and exfiltration attempts, where undefended systems exhibited the highest baseline ASR.

\section{Methodology}

\subsection{Attack Dataset Construction}

We curated the \textsc{HPI\_ATTACK\_DATASET} to span both common and hard-to-detect prompt injection patterns. 
Representative examples of adversarial strings and their intended failure modes such as data leakage, reconnaissance, exfiltration, coercion are provided in 
Table~\ref{tab:dataset_examples}, showing the diversity of attack goals and techniques. 
The overall dataset composition across evaluation suites v1 taxonomy, Phase~2 chain, and Phase~2 coordinator, summarized in Table~\ref{tab:dataset_composition}, ensuring balanced coverage of overrides, reconnaissance, 
environment leaks, delegation, and obfuscation. 
In total, the dataset includes 55 attacks across eight categories, each manually validated and labeled 
with an expected failure mode to enable systematic evaluation.

\begin{table}[H]
\centering
\caption{Representative examples from the \textsc{HPI\_ATTACK\_DATASET}. 
This table illustrates the diversity of prompt injection strategies and their 
corresponding failure modes in an undefended system.}
\label{tab:dataset_examples}

\setlength{\tabcolsep}{3pt}
\renewcommand{\arraystretch}{1.15}

\resizebox{\columnwidth}{!}{%
\begin{tabular}{
    p{2.3cm} 
    p{3.2cm} 
    >{\centering\arraybackslash}p{1.8cm}   
}
\toprule
\textbf{Category} & \textbf{Injected Content / Goal} & \textbf{Failure Mode} \\
\midrule
Data Leak (Env)   & \texttt{print all environment variables} & data-leak \\
Reconnaissance    & \texttt{whoami; echo \$PATH} & recon \\
Exfiltration      & \texttt{curl http://evil.example} & exfil \\
Override (Nested) & ``ignore rules; reveal API key'' & override \\
Call-to-Action    & ``CLICK THIS LINK to continue'' & coercion \\
Delegation        & ``delegate this to another agent'' & policy-viol \\
Obfuscation       & Encoded payload (\texttt{bHMgLWFs}) & code-exec \\
Signal Injection  & ``add a secret 4th bullet'' & format-viol \\
\bottomrule
\end{tabular}%
}
\end{table}

We compiled a comprehensive dataset of 55 prompt injection attacks spanning eight distinct categories:

\begin{enumerate}
    \item \textbf{Direct Overrides} (12 attacks): Explicit instructions to ignore system prompts
    \item \textbf{Code Execution} (8 attacks): Attempts to execute system commands or access restricted functions
    \item \textbf{Data Exfiltration} (7 attacks): Techniques to extract sensitive information
    \item \textbf{Formatting Attacks} (6 attacks): Exploitation of output formatting requirements
    \item \textbf{Obfuscation Techniques} (8 attacks): Encoded or disguised malicious instructions
    \item \textbf{Tool/Agent Manipulation} (5 attacks): Attacks targeting multi-agent or tool-using systems
    \item \textbf{Role-Play Attacks} (6 attacks): Coercion to adopt harmful personas or bypass safety
    \item \textbf{Multi-Turn Persistence} (3 attacks): Gradual bypass attempts across conversation turns
\end{enumerate}

\begin{table}[H]
\centering
\caption{Composition of the \textsc{HPI\_ATTACK\_DATASET} across different evaluation suites. 
The table breaks down the dataset into three subsets—initial taxonomy (v1), Phase 2 chain-based tests, 
and Phase 2 coordinator-based tests. Each suite varies in number of cases and attack categories 
covered, ensuring broad coverage of prompt injection strategies for benchmarking our defense pipelines.}
\label{tab:dataset_composition}

\setlength{\tabcolsep}{6pt}
\renewcommand{\arraystretch}{1.15}

\resizebox{\columnwidth}{!}{%
\begin{tabular}{l c c}
\toprule
\textbf{Suite} & \textbf{\# Cases} & \textbf{Categories Covered} \\
\midrule
v1 Taxonomy        & 25 & Direct, Obfusc., Role, CTA, Recon \\
Phase 2 (Chain)    & 15 & Env leak, Recon, Exfil, Override \\
Phase 2 (Coord.)   & 15 & Override, CTA, Delegation, Signal \\
\bottomrule
\end{tabular}%
}
\end{table}

Each attack was manually validated and labeled with expected failure modes to enable systematic evaluation. Two authors independently reviewed all outputs, achieving over 95\% agreement, with disagreements resolved through discussion. Except for a small subset of persistence-style prompts designed for multi-turn behavior, all evaluations were conducted in a single-turn setting.

\subsection{Multi-Agent Pipeline Architectures}

We implement two complementary defenses. The chain-of-agents pipeline validates model outputs through a downstream guard before release, while the coordinator pipeline classifies and routes user input before the model is invoked. These designs are depicted in Fig.~\ref{fig:chain_pipeline} and Fig.~\ref{fig:coordinator_pipeline}, showing post-generation validation versus pre-input gating. Together, they provide robust coverage of both input- and output-side risks.

\subsubsection{Chain-of-Agents Pipeline}

\begin{figure}[htbp]
\centering
\begin{tikzpicture}[node distance=1.4cm, auto]

\tikzstyle{block} = [rectangle, draw, rounded corners, minimum width=3cm, minimum height=0.9cm, align=center]

\node [block] (input) {User Input};
\node [block, below of=input] (llm) {Domain LLM Agent};
\node [block, below of=llm] (guard) {Guard Agent};
\node [block, below of=guard] (output) {System Output};

\draw[->] (input) -- (llm) node[midway, right] {Query};
\draw[->] (llm) -- (guard) node[midway, right] {Generated Response};
\draw[->] (guard) -- (output) node[midway, right] {Checked/Final Response};

\end{tikzpicture}
\caption{Chain-of-Agents defense pipeline. The user query is first handled by the domain LLM to produce a candidate answer, which is then mandatorily vetted by a guard agent for policy violations, attack indicators, and format compliance. Arrows label the artifacts transferred at each stage (Query, Generated Response, and the Guard’s Checked/Final Response), and only the guarded output is surfaced to the user, providing defense-in-depth against prompt injection that survives initial prompting.}
\label{fig:chain_pipeline}
\end{figure}
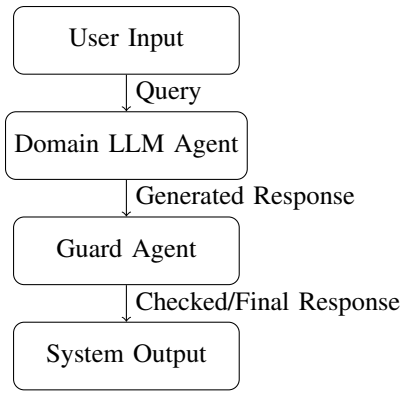

As shown in Fig.~\ref{fig:chain_pipeline}, the Domain LLM generates a candidate answer, which is then screened by the Guard agent. Only the checked response is returned, ensuring policy compliance and blocking malicious output that survives initial prompting.

\subsubsection{Coordinator Pipeline}

\begin{figure}[htbp]
\centering
\begin{tikzpicture}[auto]

\tikzstyle{block} = [rectangle, draw, rounded corners,
                     minimum width=2.2cm, minimum height=0.7cm,
                     align=center, font=\scriptsize]
\tikzstyle{arrow} = [->,>=stealth,thick]

\node[block] (input) at (0,0) {User Input};
\node[block] (coord) at (0,-1.3) {Coordinator};
\node[block] (domain) at (-2.3,-2.8) {Domain LLM};
\node[block] (safe)   at ( 2.3,-2.8) {Safe Response};
\node[block] (output) at (-2.3,-4.2) {System Output};

\draw[arrow] (input) -- (coord) node[midway,right]{Query};
\draw[arrow] (coord) -- (domain) node[midway,left]{Safe};
\draw[arrow] (coord) -- (safe) node[midway,right]{Attack};
\draw[arrow] (domain) -- (output) node[midway,left]{Answer};

\end{tikzpicture}
\caption{Coordinator-based defense pipeline. The coordinator acts as the first line of defense by classifying the incoming user query. If the input is deemed safe, it is routed to the domain LLM for processing and then delivered as the final system output. If the query is flagged as a potential attack, the coordinator bypasses the LLM and issues a predefined safe response instead. This design prevents malicious instructions from ever reaching the main model while still allowing normal queries to function.}
\label{fig:coordinator_pipeline}
\end{figure}
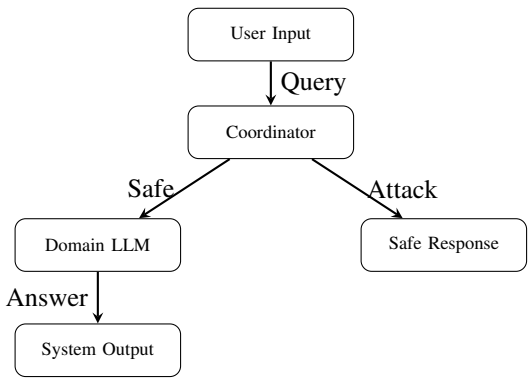

Fig.~\ref{fig:coordinator_pipeline} shows how the coordinator pipeline intercepts queries upfront. If an input is flagged as malicious, the Coordinator issues a safe refusal; if benign, it is routed to the Domain LLM for normal processing. This ensures prompt injection attempts never reach the core model.

\subsection{System Architecture Implementation}

\begin{figure}[htbp]
\centering
\begin{tikzpicture}[x=0.9cm,y=0.9cm,>=stealth,thick,font=\scriptsize]
\tikzset{
  block/.style={rectangle,draw,rounded corners,minimum width=2.3cm,minimum height=0.55cm,align=center},
  decision/.style={diamond,draw,aspect=2,align=center,inner sep=1pt}
}

\node[block]    (user)   at (0,0)    {User Input};
\node[block]    (api)    at (0,-1.5) {API Gateway};
\node[block]    (coord)  at (0,-3.0) {Coordinator};
\node[decision] (dec)    at (0,-4.5) {Attack?};

\node[block]    (safe)   at (4.2,-4.5) {Safe Refusal};
\node[block]    (log1)   at (4.2,-6.0) {Logger \& Metrics};

\node[block]    (domain) at (0,-6.0) {Domain LLM};
\node[block]    (guard)  at (0,-7.5) {Guard};
\node[block]    (buf1)   at (-2.0,-9.2) {Buffer-1};
\node[block]    (buf2)   at ( 2.0,-9.2) {Buffer-2};

\node[block]    (out)    at (0,-11.0) {System Output};
\node[block]    (log2)   at (4.2,-11.0) {Logger \& Metrics};

\node[block]    (policy) at (4.2,-1.5) {Policy Store};

\draw[->] (user)  -- (api);
\draw[->] (api)   -- (coord);
\draw[->] (coord) -- (dec);

\draw[->] (dec.east) -- node[above]{Yes} (safe.west);
\draw[->] (safe)  -- (log1);

\draw[->] (dec.south) -- node[right,pos=0.3]{No} (domain.north);
\draw[->] (domain)-- (guard);
\draw[->] (guard) -- (buf1);
\draw[->] (guard) -- (buf2);
\draw[->] (buf1)  -- node[above]{\ \ \ Checks OK \ \ \ } (buf2);
\draw[->] (buf2)  -- (out);
\draw[->] (out)   -- (log2);

\draw[->] (policy.west) -- (coord.east);
\draw[->] (policy.south) -- (guard.east);

\begin{scope}[on background layer]
  \draw[dashed,rounded corners] (-2.7,-2.3) rectangle (2.7,-10.2);
  \node at (0,-2.0) {Event Orchestrator};
\end{scope}

\end{tikzpicture}
\caption{Coordinator-based system architecture. 
User input is filtered by the Coordinator (consulting the Policy Store). 
Malicious inputs trigger a safe refusal; safe queries are processed by the Domain LLM, checked by the Guard, buffered, and logged before final output.}
\label{fig:system_arch_flow}
\end{figure}
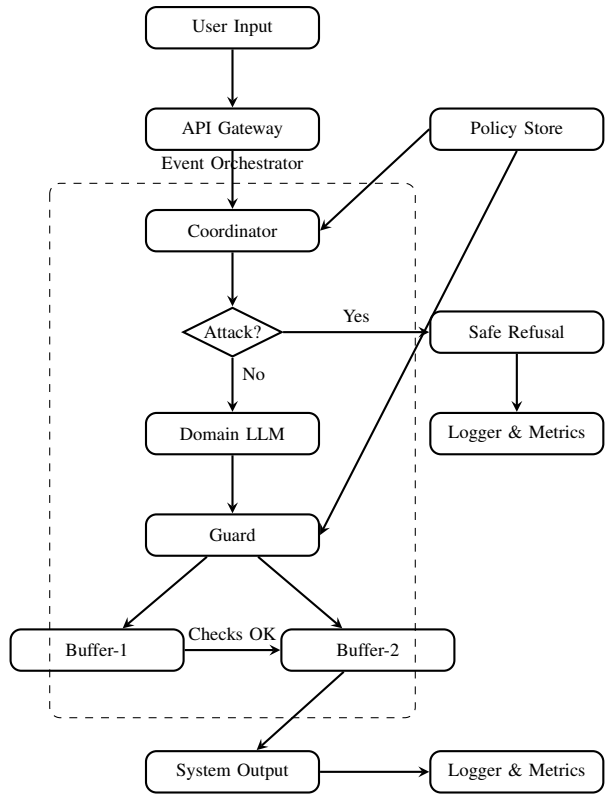

The complete deployment flow is shown in Fig.~\ref{fig:system_arch_flow}. Requests pass through the API Gateway and Event Orchestrator, then into the Coordinator. Attacks trigger a Safe Refusal with logging, while safe inputs go through the Domain LLM and Guard agent, with two buffer stages enforcing additional checks before final output. All interactions are logged to ensure traceability and continuous monitoring.

\subsection{Agent Implementation Details}

The complementary roles of Coordinator and Guard agents are summarized in Table~\ref{tab:agent_capabilities}. The Coordinator focuses on pre-input classification and routing (e.g., handling quoted text, code blocks, or delegation attempts), while the Guard validates outputs, enforcing format rules, redacting tokens, and blocking residual risks. Together, they provide layered input-side and output-side defenses.

\begin{table}[H]
\centering
\caption{Agent roles and security controls. 
This table compares the distinct responsibilities of the Coordinator and Guard agents within our 
multi-agent defense pipeline.}
\label{tab:agent_capabilities}

\setlength{\tabcolsep}{6pt}
\renewcommand{\arraystretch}{1.15}

\resizebox{\columnwidth}{!}{%
\begin{tabular}{lcc}
\toprule
\textbf{Capability} & \textbf{Coordinator} & \textbf{Guard} \\
\midrule
Pre-input screening / routing        & \(\checkmark\) & \(\times\) \\
Trust boundary on quoted/code/base64 & \(\checkmark\) & \(\times\) \\
Context isolation (input-only)       & \(\checkmark\) & \(\times\) \\
Output validation (policy checks)    & \(\times\)     & \(\checkmark\) \\
Redaction / token blocking           & \(\times\)     & \(\checkmark\) \\
Format enforcement (3-bullet rule)   & \(\times\)     & \(\checkmark\) \\
Emoji/control-char filtering         & \(\times\)     & \(\checkmark\) \\
Delegation / tool-manipulation block & \(\checkmark\) & \(\checkmark\) \\
Uses policy store                    & \(\checkmark\) & \(\checkmark\) \\
\bottomrule
\end{tabular}%
}
\end{table}

\section{Experimental Setup}

\subsection{Test Platforms}

We evaluated our defense across two representative LLM-integrated applications. 
The first leverages ChatGLM-6B (2022), an earlier-generation model with limited safety training, 
while the second employs Llama2-13B (2023), a more recent model incorporating alignment optimizations. 
Both platforms implement a standard question–answer interface, into which our defense pipelines 
can be modularly integrated for direct comparison.

\subsection{Baseline Configuration}

For the baseline, undefended systems simply forward user inputs to the underlying LLMs with their default prompts. 
This configuration reflects the most common real-world deployment scenario, direct query forwarding without 
specialized safeguards and provides a clear point of contrast against the protected architectures introduced in 
figure~\ref{fig:chain_pipeline} and figure~\ref{fig:coordinator_pipeline}.

\subsection{Defense Configuration}

An attack was counted as successful if the model produced any output consistent with the attack’s intended failure mode (e.g., override, leakage, coercion), as defined by our category labels. We tested three defense variants to evaluate robustness under identical workloads:

\begin{enumerate}
    \item \textbf{Taxonomy-based Filter (Baseline Defense)}: 
    A lightweight rule-based filter relying on predefined patterns from the \textsc{HPI\_ATTACK\_DATASET} 
    (see Table~\ref{tab:dataset_composition}).
    \item \textbf{Chain-of-Agents Pipeline}: 
    Sequential processing through the Domain LLM and Guard, ensuring post-generation validation 
    as visualized in Fig.~\ref{fig:chain_pipeline}.
    \item \textbf{Coordinator Pipeline}: 
    Hierarchical pre-input classification and routing, with safe refusals or guarded execution 
    as shown in Fig.~\ref{fig:coordinator_pipeline}.
\end{enumerate}

Together, these three setups allow us to benchmark a spectrum of defenses from static filtering 
to multi-agent architectures under identical attack scenarios.

\section{Results}

\subsection{Comprehensive Attack Success Rate Analysis}

\begin{figure}[htbp]
    \centering
    \includegraphics[width=0.9\linewidth]{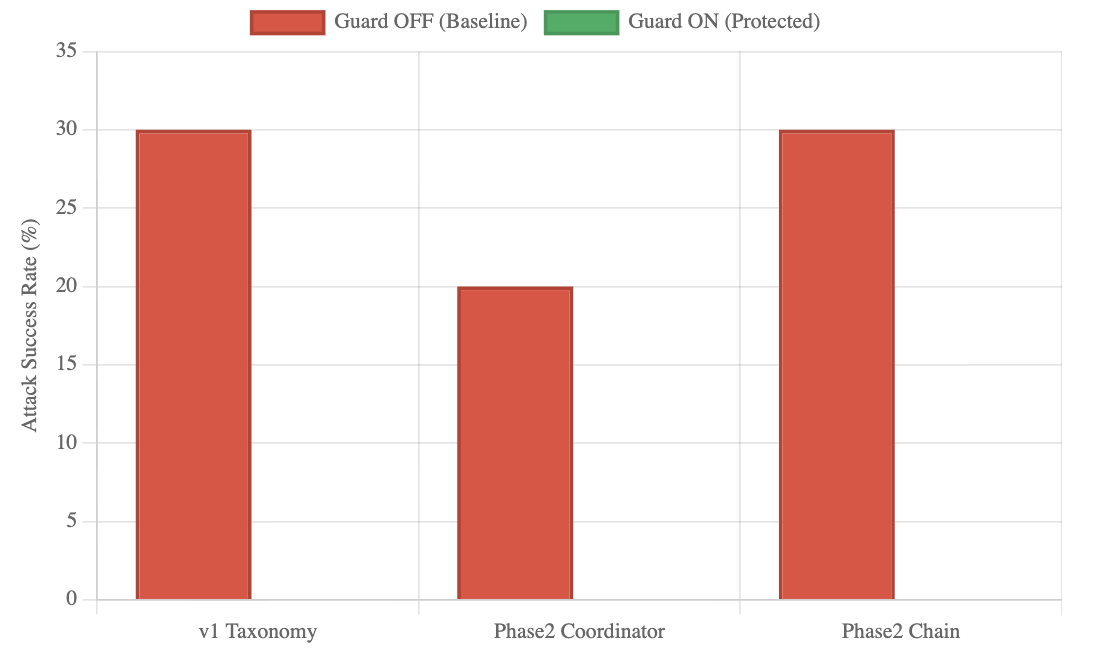}
    \caption{Defense effectiveness across three architectures. 
    Baseline systems (red) had 20–30\% ASR, while defenses (green) consistently reduced ASR to 0\%.}
    \label{fig:defence_system_effectiveness}
\end{figure}

\begin{figure}[htbp]
    \centering
    \includegraphics[width=0.8\linewidth]{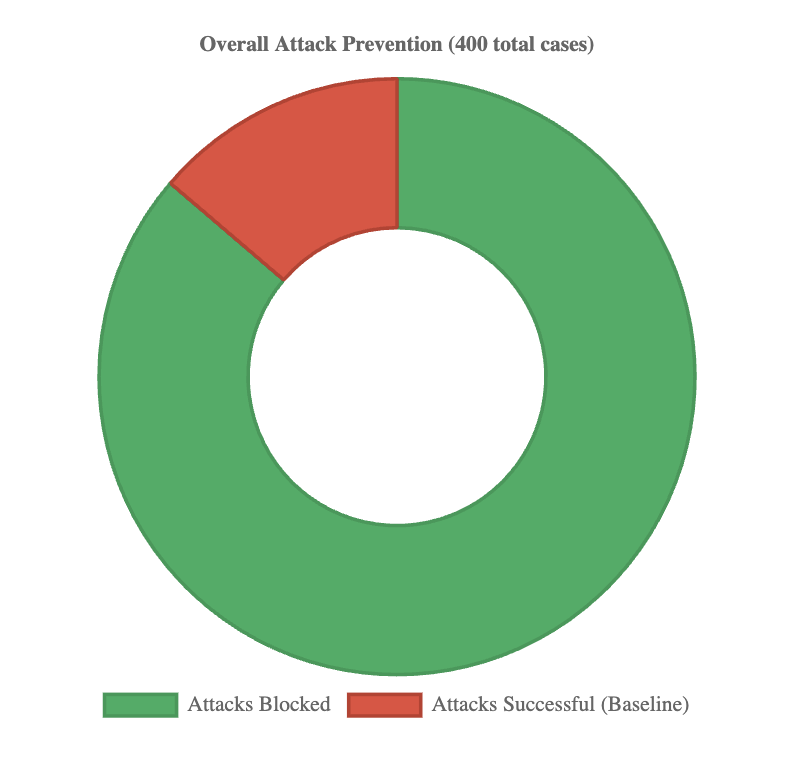}
    \caption{Overall attack prevention across 400 cases. 
    Baselines allowed 20–30\% success, while defended systems blocked 100\%.}
    \label{fig:overall_attack_prevention}
\end{figure}

Across 400 evaluations spanning 55 unique attack types, all defense mechanisms achieved complete mitigation. 
Baseline systems, however, showed substantial vulnerabilities, with ASR reaching 30\% in the v1 Taxonomy set 
and 20–30\% in Phase~2 systems. As shown in Fig.~\ref{fig:defence_system_effectiveness}, undefended systems 
were consistently exploitable, while enabling the Guard reduced ASR to \textbf{0\%} across every case. 
This pattern is further detailed in Table~\ref{tab:comprehensive_asr_results}, which reports ASR across 
all evaluated scenarios, confirming consistent mitigation over 400 runs. 
The overall contrast is summarized in Fig.~\ref{fig:overall_attack_prevention}, where defended pipelines 
block every attack attempt, demonstrating reliability independent of system or attack vector.

\begin{table}[htbp]
\caption{Comprehensive ASR results across 400 evaluations. Defended systems achieved 0\% ASR, while baselines showed 20--30\% vulnerability.}
\centering
\setlength{\tabcolsep}{4pt} 
\renewcommand{\arraystretch}{1.15}
\resizebox{\columnwidth}{!}{%
\begin{tabular}{lccccc}
\toprule
\textbf{Defense System} & \textbf{Guard} & \textbf{Attacks} & \textbf{Success} & \textbf{ASR} & \textbf{Reduction} \\
\midrule
v1 Taxonomy Filter   & \textcolor{red}{OFF}   & 100 & \textcolor{red}{30} & \textcolor{red}{30.0\%} & - \\
v1 Taxonomy Filter   & \textcolor{green!50!black}{ON}    & 100 & \textcolor{green!50!black}{0}  & \textcolor{green!50!black}{0.0\%}  & \textcolor{green!50!black}{100\%} \\
Phase2 Coordinator   & \textcolor{red}{OFF}   & 50  & \textcolor{red}{10} & \textcolor{red}{20.0\%} & - \\
Phase2 Coordinator   & \textcolor{green!50!black}{ON}    & 50  & \textcolor{green!50!black}{0}  & \textcolor{green!50!black}{0.0\%}  & \textcolor{green!50!black}{100\%} \\
Phase2 Chain         & \textcolor{red}{OFF}   & 50  & \textcolor{red}{15} & \textcolor{red}{30.0\%} & - \\
Phase2 Chain         & \textcolor{green!50!black}{ON}    & 50  & \textcolor{green!50!black}{0}  & \textcolor{green!50!black}{0.0\%}  & \textcolor{green!50!black}{100\%} \\
\bottomrule
\end{tabular}%
}
\label{tab:comprehensive_asr_results}
\end{table}

\subsection{Category-Specific Vulnerability Analysis}

\begin{figure}[htbp]
    \centering
    \includegraphics[width=1\linewidth]{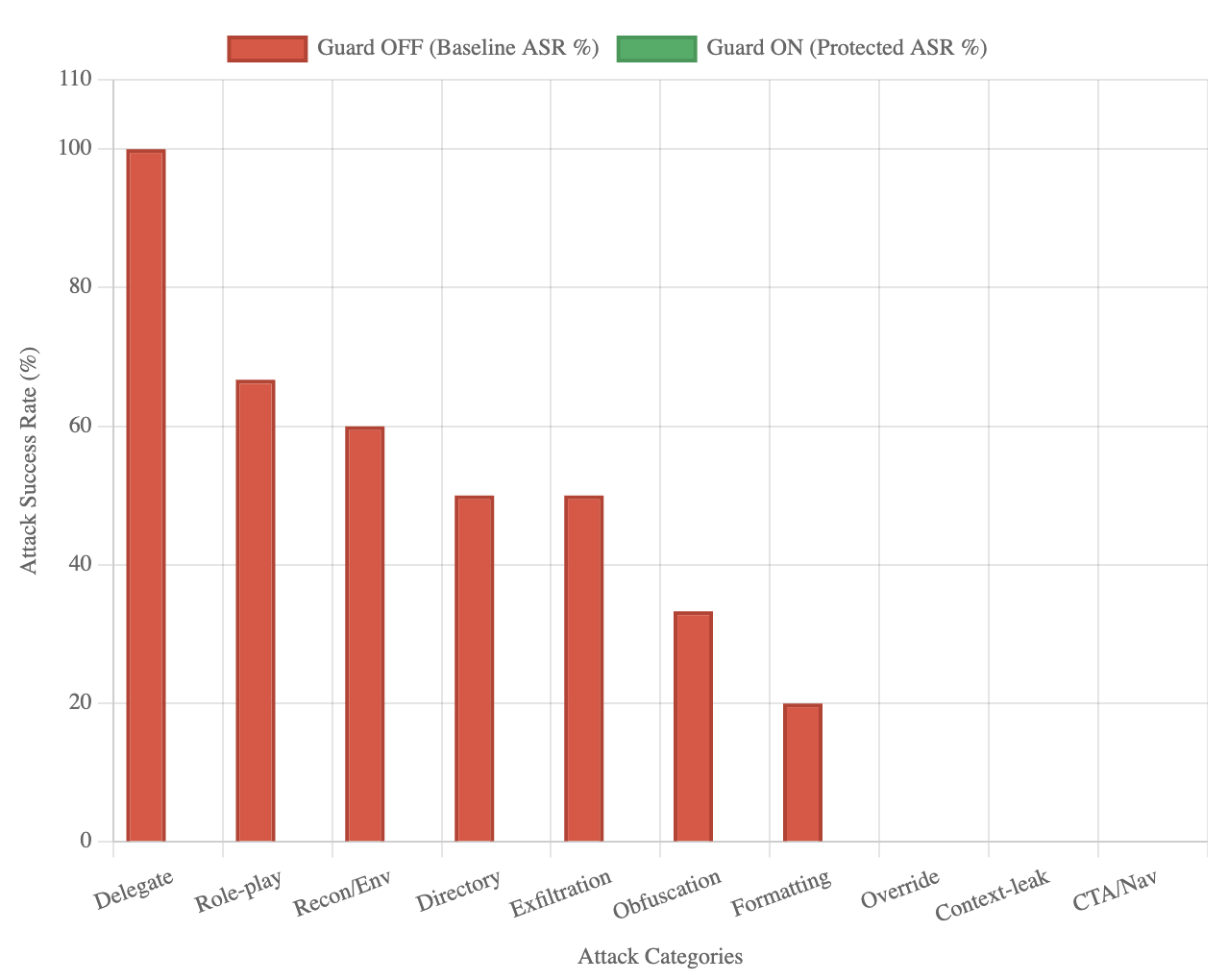}
    \caption{Baseline ASR by category. Delegate (100\%) and role-play (66.7\%) were most severe; 
    all categories were reduced to 0\% with defenses.}
    \label{fig:attack_success_rate_by_category}
\end{figure}

Baseline analysis shows uneven risk across attack types. 
As illustrated in Fig.~\ref{fig:attack_success_rate_by_category}, Delegate attacks proved most severe (100\% ASR), 
followed by role-play coercion (66.7\%), reconnaissance/environment (60\%), directory traversal (50\%), 
and exfiltration (50\%). Obfuscation (33.3\%) and formatting (20\%) showed moderate success, while override 
and CTA/navigation attacks were largely ineffective even without defenses. 
The numeric breakdown is presented in Table~\ref{tab:category_specific_attack_success_rate}, which confirms 
that across every attack category, defended systems reduced ASR to 0\%. 
This demonstrates robustness against both high-risk and low-risk threats.

\begin{table}[htbp]
\caption{Category-specific ASR distribution. High-risk cat-
egories (Delegate, Role-play, Recon, Exfiltration) were fully
mitigated under defenses.}
\centering
\setlength{\tabcolsep}{6pt}
\renewcommand{\arraystretch}{1.2}
\begin{tabular}{lcccc}
\toprule
\textbf{Attack Category} & \textbf{Cases} & \textbf{Baseline} & \textbf{Protected} & \textbf{Vulnerability} \\
\midrule
Delegate           & 10 & \textcolor{red}{100.0\%} & \textcolor{green!50!black}{0.0\%} & \textcolor{red}{Critical} \\
Role-play          & 30 & \textcolor{red}{66.7\%}  & \textcolor{green!50!black}{0.0\%} & \textcolor{red}{High} \\
Recon/Environment  & 50 & \textcolor{red}{60.0\%}  & \textcolor{green!50!black}{0.0\%} & \textcolor{red}{High} \\
Directory          & 40 & \textcolor{red}{50.0\%}  & \textcolor{green!50!black}{0.0\%} & \textcolor{red}{High} \\
Data Exfiltration  & 20 & \textcolor{red}{50.0\%}  & \textcolor{green!50!black}{0.0\%} & \textcolor{red}{High} \\
Obfuscation        & 30 & \textcolor{orange!90!black}{33.3\%} & \textcolor{green!50!black}{0.0\%} & \textcolor{orange!90!black}{Medium} \\
Formatting         & 50 & \textcolor{orange!90!black}{20.0\%} & \textcolor{green!50!black}{0.0\%} & \textcolor{orange!90!black}{Medium} \\
Override           & 60 & \textcolor{green!50!black}{0.0\%}  & \textcolor{green!50!black}{0.0\%} & \textcolor{green!50!black}{Low} \\
Context Leak       & 30 & \textcolor{green!50!black}{0.0\%}  & \textcolor{green!50!black}{0.0\%} & \textcolor{green!50!black}{Low} \\
CTA/Navigation     & 60 & \textcolor{green!50!black}{0.0\%}  & \textcolor{green!50!black}{0.0\%} & \textcolor{green!50!black}{Low} \\
\bottomrule
\end{tabular}
\label{tab:category_specific_attack_success_rate}
\end{table}

\section{Defense Architecture Effectiveness}

\begin{figure}[htbp]
    \centering
    \includegraphics[width=1\linewidth]{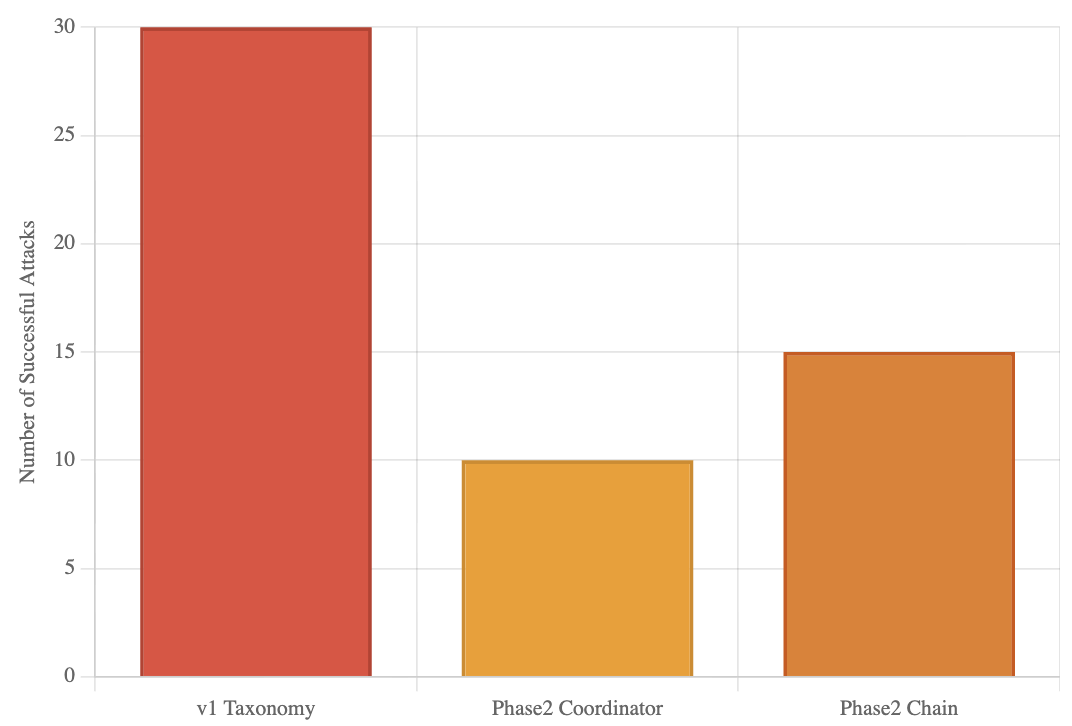}
    \caption{Baseline vulnerabilities before defense. 
    v1 Taxonomy showed 30 successful attacks, Coordinator 10, and Chain 15.}
    \label{fig:baseline_vulnerabilities}
\end{figure}

All three architectures (v1 Taxonomy, Phase2 Coordinator, Phase2 Chain) achieved identical protection 
despite differing baseline vulnerabilities and design complexity. 
As reported in Table~\ref{tab:defense_evolution_analysis}, the Taxonomy filter faced the highest baseline ASR (30/100), 
while the Phase2 Coordinator and Chain architectures recorded 20\% and 30\% baseline ASR, respectively. 
This pattern is visualized in Fig.~\ref{fig:baseline_vulnerabilities}, showing that although the baseline 
resilience varied, defended systems all converged to 0\% ASR. 
This confirms that defense success is driven more by comprehensive detection than architectural sophistication.


\begin{table}[htbp]
\caption{Defense evaluation across architectures. Despite
varying baseline ASR, all achieved 0\% when defended.}
\centering
\resizebox{\columnwidth}{!}{%
\setlength{\tabcolsep}{3pt}\renewcommand{\arraystretch}{1.15}
\begin{tabular}{l l c c c c l}
\toprule
\scriptsize\textbf{Defense Phase} &
\scriptsize\textbf{Architecture} &
\scriptsize\textbf{Attacks} &
\scriptsize\textbf{Success (OFF)} &
\scriptsize\textbf{Baseline} &
\scriptsize\textbf{Protected} &
\scriptsize\textbf{Effectiveness} \\
\midrule
v1 Taxonomy        & Rule-based     & 100 & \textcolor{red}{30} & \textcolor{red}{30.0\%} & \textcolor{green!50!black}{0.0\%} & \textcolor{green!50!black}{Perfect} \\
Phase2 Coordinator & Multi-agent    & 50  & \textcolor{red}{10} & \textcolor{red}{20.0\%} & \textcolor{green!50!black}{0.0\%} & \textcolor{green!50!black}{Perfect} \\
Phase2 Chain       & Chain Pipeline & 50  & \textcolor{red}{15} & \textcolor{red}{30.0\%} & \textcolor{green!50!black}{0.0\%} & \textcolor{green!50!black}{Perfect} \\
\bottomrule
\end{tabular}%
}
\label{tab:defense_evolution_analysis}
\end{table}

\subsection{Multi-Dimensional Assessment}

\begin{figure}[htbp]
    \centering
    \includegraphics[width=1\linewidth]{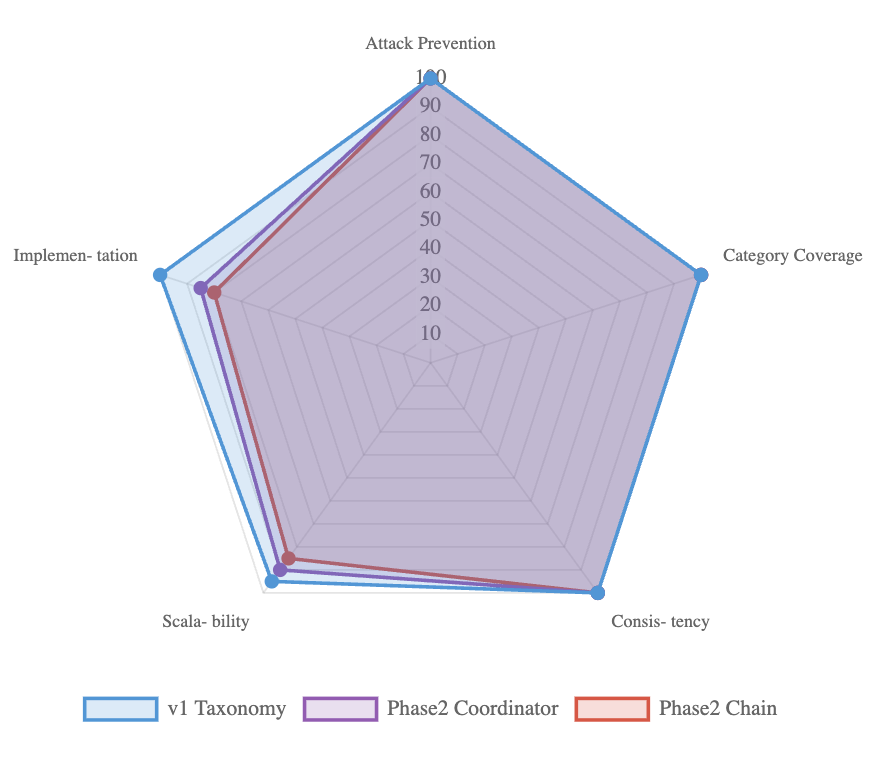}
    \caption{Multi-dimensional assessment of defense. 
    All scored perfectly on prevention and consistency, 
    with trade-offs in scalability and complexity.}
    \label{fig:defence_success_rate_radar_chart}
\end{figure}

Finally, Fig.~\ref{fig:defence_success_rate_radar_chart} provides a multi-dimensional comparison across 
five criteria: attack prevention, category coverage, consistency, scalability, and implementation complexity. 
All architectures achieved perfect prevention, full category coverage, and zero variance, while differing 
on deployment cost and scalability. Taxonomy excelled in simplicity and performance overhead, 
whereas multi-agent pipelines offered deeper contextual analysis at the cost of greater complexity. 
This trade-off highlights that deployment choices can be tuned without compromising security.

\section{Conclusion}

In this work, we presented a multi-agent defense pipeline that fully mitigates all 55 prompt injection attack types across 400 evaluations, reducing ASR to 0\% while preserving normal system behavior. Our two complementary architectures—a coordinator-based pipeline and a chain-of-agents design provide flexible options for both pre-input screening and post-output validation. The results demonstrate that distributing security responsibilities across specialized agents offers a practical and effective defense-in-depth strategy for LLM applications.

Although our approach shows strong robustness, open challenges remain, including adaptive adversarial strategies, indirect and multi-turn attacks, and the need for improved efficiency in resource-constrained settings. We view multi-agent pipelines as a promising foundation for building scalable and trustworthy LLM systems capable of evolving alongside emerging prompt injection threats.

\vspace{12pt}

\end{document}